\documentclass[aps,twocolumn,preprintnumbers,amsmath,amssymb,footinbib,prl]{revtex4}
\usepackage{graphicx}
\usepackage{bm}
\newcommand{\be}{\begin{equation}}
\newcommand{\ee}{\end{equation}}
\newcommand{\bea}{\begin{eqnarray}}
\newcommand{\eea}{\end{eqnarray}}

\begin{document}
\title{Inhomogeneities in the freeze-out of relativistic heavy ion
collisions at CERN SPS}
\author{Detlef Zschiesche}
\affiliation{
Institut f\"ur Theoretische Physik, J.~W.~Goethe Universit\"at,\\
Max-von-Laue-Strasse 1,
D-60439 Frankfurt am Main, Germany\\
}

\begin{abstract}
We study the role of temperature and density inhomogeneities 
on the freeze-out of relativistic heavy ion collisions at CERN SPS. 
Especially the impact on the 
particle abundancies is investigated. The quality of the 
fits to the measured
particle ratios in 158 AGeV Pb+Pb collisions 
significantly improves as compared to a homogeneous model. 
\end{abstract}
\maketitle
\begin{figure}[h]
\includegraphics[width=8cm]{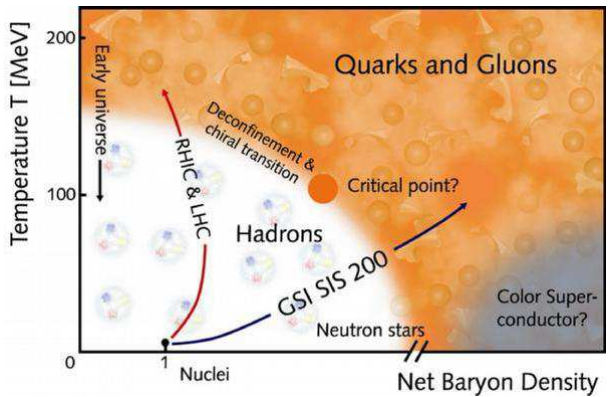}
\vspace{-0.2cm}
\caption{\label{qcdpd}
Different phases of strongly interacting matter in the 
T-$\rho$ plane. Relativistic heavy ion
experiments heat and compress ordinary nuclei to high temperatures and
densities.  
Depending on the bombarding energy, the transition 
to chirally restored quark- and gluon matter happens in different 
temperature and density regions, where the characteristics of the 
phase transition are expected to be different.
Taken from \protect{\cite{GSI}}.}
\end{figure}

\noindent
One of the key motivations for the    
heavy ion programs at GSI, CERN and BNL 
is to shed light on the 
QCD phase diagram. More specifically,   
the aim is to gain a deeper 
understanding of  
the physics of the different
phases of QCD matter and of the characteristics of the deconfinement and 
chiral phase transition (cf. Fig. \ref{qcdpd}). 

The current picture of the QCD phase diagram is as follows: 
At vanishing chemical potential ($\mu=0$) finite temperature lattice QCD 
calculations find a rapid but smooth crossover 
(see e.g. \cite{Brown90}). At large $\mu$  
one has to rely on model calculations, since lattice
QCD calculations encounter the fermion sign problem. 
However, several different model
calculations (see \cite{Stephanov04} for a summary)
suggest a first order phase
transition. Combining these two results, 
the line of first order phase transitions originating 
at the $T=0$ axis cannot end at $\mu=0$ but at some 
point in the  ($T_c, \mu_c$) plane with finite $\mu$. 
At this endpoint a second order phase transition is expected.
For chemical potentials smaller than $\mu_c$ a crossover occurs. 
This picture is also
supported by different extrapolations of lattice QCD to finite
chemical potential \cite{Allton02,Fodor:2001pe}.

\begin{figure}[h]
\vspace{-1cm}
\includegraphics[width=9cm]{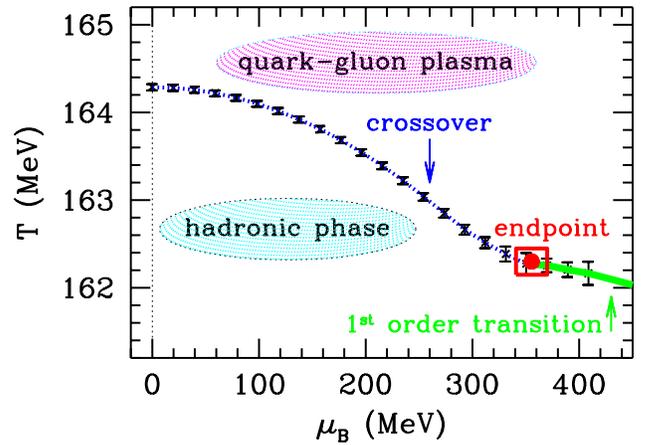}
\vspace{-1.8cm}
\caption{\label{fodor}
Quark-hadron phase transition as obtained from lattice calculations 
(at finite chemical potential). Taken from {\protect{\cite{Fodor:2001pe}}}.}
\end{figure}

E.g., in \cite{Fodor:2001pe} it was found that
a line of first-order phase transitions in
the $(\mu_B,T)$ plane ends in a critical point at
$T\approx160$~MeV, $\mu_B\approx360$~MeV (cf. Fig. \ref{fodor}).

Heavy-ion collisions at high energies are hoped to be able 
to detect that critical point
and verify this picture. 
For high enough energies the
phase transition/crossover line should be crossed - the critical
energy density is expected to be reached at intermediate SPS energies
or the new GSI facility. Since the system takes different 
paths in the $T-\mu$ or $T-\rho$ plane for different bombarding 
energies, it is hoped that by varying the
beam energy, one can ``switch'' between the regimes of first-order
transition and cross over, respectively (cf. Fig. \ref{qcdpd}). 

But how do we know whether the system passed through 
a first order phase transition, a
crossover, or if no phase change at all occured?

Using a non-equilibrium hydrodynamical simulation it was shown in  
\cite{Paech:2003fe} that  the expanding 
fluid develops significant inhomogeneities, if a first order
phase transition is crossed. These inhomogeneities should also be
present on the decoupling surface of the hadrons.
\begin{figure}[h]
\includegraphics[width=9cm]{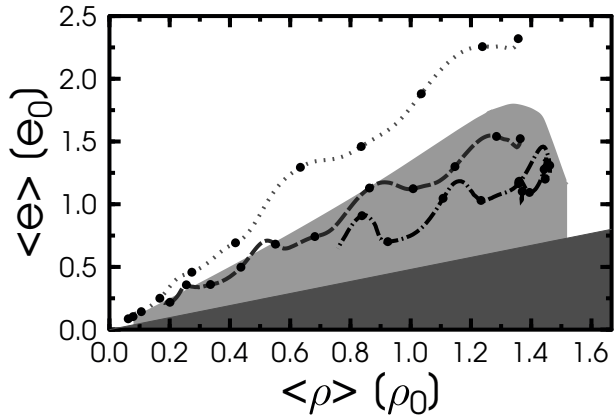}
\vspace{-0.8cm}
\caption{\label{col1}
Evolution of the average fluid energy
through a crossover (dots), and a weak (dashes) and strong (dash-dots)
first order phase transition.
The fat dots indicate time intervals of $\approx 1.5$ fm/c. 
Taken from \cite{Paech:2003fe}.}
\end{figure}

\begin{figure}[h]
\includegraphics[width=8.5cm]{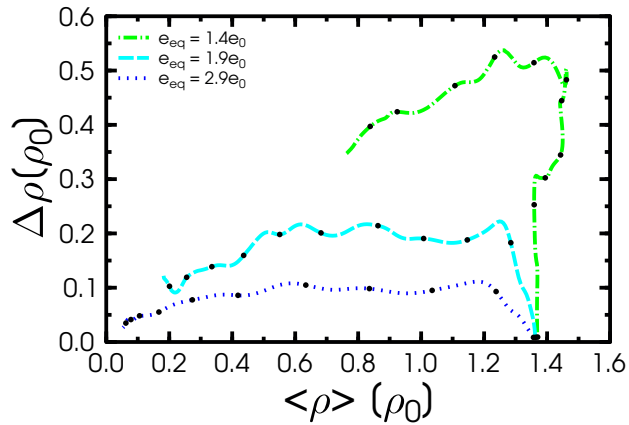}
\vspace{-0.5cm}
\caption{\label{col2} RMS fluctuation of the baryon density with
initial conditions chosen such that the system undergoes a crossover
(dots), weak (dashes) and strong (dash-dots) first order transition as
a function of the average baryon density. The fat dots indicate time
intervals of $\approx 1.5$ fm/c. Taken from \cite{Paech:2003fe}.}
\end{figure}

Fig. \ref{col1} shows the trajectory of the system within the phase
diagram for different initial conditions. Depending on 
these initial conditions, the system either evolves smoothly through a
crossover or enters the region corresponding to phase coexistence 
in the equilibrium phase diagram and thus undergoes a first order
phase transition. The resulting RMS fluctuations of the baryon density
are shown in figure \ref{col2}. It can be seen that the amplitude of
the density contrast is substantially larger for a strong first order
transition (initial energy density $\rm{e}_{\rm{eq}}=1.4 e_0$) than for a crossover ($\rm{e}_{\rm{eq}}=2.9 e_0$).

\begin{figure}[h]
\includegraphics[width=8cm]{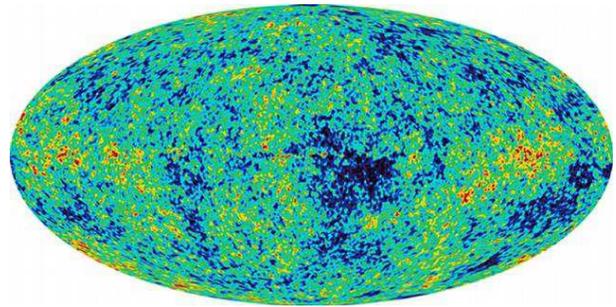}
\caption{\label{wmap}
CMB photon
decoupling surface observed by WMAP.
Dark regions correspond to cooler and brighter regions corrspond to 
warmer spots.
Taken from \protect{\cite{wmap}}.
}
\end{figure}

That means, that 
if the particles decouple shortly after the expansion trajectory
crosses the line of first order transitions one may expect
a rather inhomogeneous (energy-) density distribution on the
freeze-out surface~(similar, say, to the CMB photon
decoupling surface observed by WMAP~\cite{wmap}, see Fig. \ref{wmap}).

On the other hand, if the system expands through a crossover transition, 
as expected for 
collisions at
very high energies ($\mu_B\simeq0$, cf. Fig. \ref{qcdpd}) 
it may cool smoothly from high to low
$T$ and so pressure gradients tend to wash out density
inhomogeneities. 
Similarly, in the absence of phase-transition induced
non-equilibrium effects, the predicted initial-state density
inhomogeneities~\cite{iniflucs} should be strongly damped.

Unfortunately, if the scale of the inhomogeneities is much
smaller than the decoupling volume then
they can not be resolved individually, nor will they give rise to large
event-by-event fluctuations. However, because of the nonlinear
dependence of the hadron densities on $T$ and $\mu_B$ they should nevertheless
reflect in the {\em event-averaged} abundances.

Particle production in
relativistic heavy ion collisions has been investigated in several works 
(see for example \cite{thermo})
using 
(homogeneous) thermodynamical equilibrium calculations.
In addition, e.g. in \cite{thermo_gammas,Becattini:2003wp} 
extensions accounting for 
strange- and light quark non-equlibrium 
were considered and for example in \cite{thermo_int} the role of in-medium masses 
was discussed.
Here, we 
attempt to check whether the experimental data show any signs of
inhomogeneities on the freeze-out surface. 
To do this we investigate 
an inhomogeneous fireball at (chemical)
decoupling. 
Perhaps the simplest possible {\it ansatz} 
is to employ the grand
canonical ensemble and - in extension to the homegeneous models - 
assume that the intensive
variables $T$ and $\mu_B$ are distributed according to a
Gaussian \cite{Dumitru05}. 
This avoids reference to any
particular dynamical model for the formation and the distribution of density
perturbations on the freeze-out surface. Also, in this simple model
we do not need to
specify the probability distribution of volumes $V$.
Then, the average density of species $i$ is computed as
\bea
& &\overline{\rho}_i\; (\overline{T},\overline{\mu}_B, \delta T,\delta\mu_B)
= \\
& &\int\limits_0^\infty
 dT \; P(T;\overline{T},\delta T) 
\int\limits_{-\infty}^\infty 
d\mu_B \; P(\mu_B;
\overline{\mu}_B,\delta\mu_B)~\rho_i (T,\mu_B)~, \nonumber
\eea
with $\rho_i(T,\mu_B)$ the actual ``local'' density of species $i$, and
with $P(x; \overline{x},\delta x) \sim $ 
$\exp \left(-\frac{\left(x-\overline{x}\right)^2}{2\; \delta x^2} \right)$
the distribution of temperatures and chemical potentials on the
freeze-out surface.  
Feeding from (strong or weak) decays is included by replacing
$\overline{\rho}_i \rightarrow \overline{\rho}_i + B_{ij}\;
\overline{\rho}_j.$
The implicit sum over $j\neq i$ runs over all unstable hadron species, with
$B_{ij}$ the branching ratio for the decay $j\to i$. 
For the present analysis we computed the densities $\rho_i (T,\mu_B)$ in the ideal-gas
approximation. The resonances are included up to 1.5 GeV in mass for the mesons and
up to 2 GeV for the baryons. The finite widths of the resonances were
not taken into account and 
unknown branching ratios in the particle data book were excluded from
the feeding.  Furthermore, we use a four-dimensional table with 5 MeV
steps in $T$ and $\delta T$ and $10$ MeV steps in  
$\mu$ and $\delta \mu$. This finite grid-size of course 
limits our accuracy in determining the best fits.
However, our approach should be well suited to investigate 
the qualitative behavior of the parameters $\delta T$ and $\delta
\mu_B$ and 
whether they can significantly improve the agreement with the 
experimental data.

The data used in our analysis are the particle multiplicities 
measured by the NA49 collaboration in 
$\sqrt{s_{NN}} = 17.3$ GeV
Pb+Pb collisions at CERN SPS. 
We use midrapidity and $4 \pi$ data, both as   
compiled in \cite{Becattini:2003wp}. 

Using these data, we perform a $\chi^2$ fit.
I.e., we determine the minimal value of 
\begin{equation}
\chi^2 = \sum_i {\left(r_i^{exp} - r_i^{model}\right)^2}/{\sigma_i^2},
\end{equation}
where $r_i^{exp}$, $r_i^{model}$ denote the experimentally measured
and the calculated particle ratios, respectively, and  
$\sigma_i^2$ is the experimental error. We compare two cases: On the on hand 
the homogeneous fit, where the values of $\delta T$ and $\delta \mu$ are set 
to zero, and on the other hand the inhomogeneous fit, where we allow for finite 
values of $\delta T$ and $\delta \mu$.

We find that the fits improve (lower $\chi^2/\rm{dof}$) substantially if 
$\delta T$, $\delta\mu_B$ are not forced to
zero.  Table \ref{tab1} shows the resulting best fits, with and
without finite widths of the $T$ and $\mu_B$ distributions.

\begin{table}[ht]
\begin{small}
\begin{center}
\begin{tabular}{|c|ccccc|} \hline
 & $\overline{T}$ & $\overline{\mu}_B$ & $\delta T$ & 
      $\delta\mu_B$ & $\chi^2/\mbox{dof}$ \\ \hline
\hline
 SPS-158&$155\pm 5$ & $200\pm 10$ & 0 & 0 & 40.4/8 \\ 
(mid)& $105\pm 5$ & $230\pm 15$ & $35\pm 5$ & $80 \pm 40$ & 11.2/6 \\ 
\hline
SPS-158  & $145\pm 5$ & $210\pm 15$ & 0 & 0 & 40.0/11 \\
($4\pi$)&$100\pm 5$ & $260\pm 15$ & $30\pm5$ & $190\pm 35$ & 5.7/9 \\ 
\hline
\end{tabular}
\end{center}
\end{small}
\vspace{-0.5cm}
\caption{\label{tab1} Best fit parameters and $\chi^2/dof$ to SPS 158
AGeV midrapidity and
$4 \pi$ data, measured by the NA49 collaboration. 
The cases $\delta T = \delta \mu = 0$ corresponds to the homogeneous 
freeze-out model and those with finite values for $\delta T, \delta
\mu$ correspond to the inhomogeneous model.
}
\end{table}

As can be seen, for $4 \pi$ data the resulting best fit $\chi^2/dof$
values are
approximately 3.6 for the homogeneous fit and 0.63 for the inhomogeneous fit.
For midrapidity data we obtain $\chi^2/dof \approx 5.1$
for the homogeneous  
and $\chi^2/dof \approx 1.9$ for the inhomogeneous case. 
I.e., for both data sets the $\chi^2$ per degree of freedom is considerably
reduced by allowing for inhomogeneities or finite widths of the 
$T$- and $\mu$-distributions, respectively. In other words: 
for midrapidity as well as for $4 \pi$ data $\delta T$
and $\delta \mu$ represent significant paramters. 
Error estimates for the parameters 
(confidence intervals) are obtained from the projection of 
the regions in parameter space defined by $\chi^2 \le \chi^2_{min} +
1$ onto each axis. This corresponds to a
confidence level of $68.3 \%$ if the errors are normally distributed.
As shown in table \ref{tab1}, 
within this error estimate, the best-fit values for $\delta T$ and
$\delta \mu$ are significantly greater than 0.
 
\begin{figure}[h]
\vspace{-0.7cm}
\includegraphics[width=9.5cm]{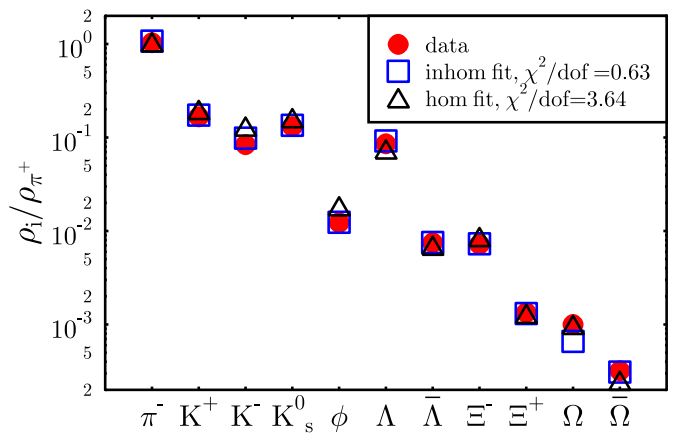}
\vspace{-0.6cm}
\caption{\label{ratios_log}
Particle ratios ($4 \pi$, compiled in
\protect{\cite{Becattini:2003wp}}) 
as measured by the NA49 collaboration in SPS 158 AGeV 
Pb+Pb collisions compared to the homogeneous fit 
($\delta T = \delta \mu = 0$) and the inhomogeneous fit 
($\delta T, \delta \mu$ free parameters). 
}
\end{figure}

\begin{figure}[h]
\vspace{-1.0cm}
\includegraphics[width=9cm]{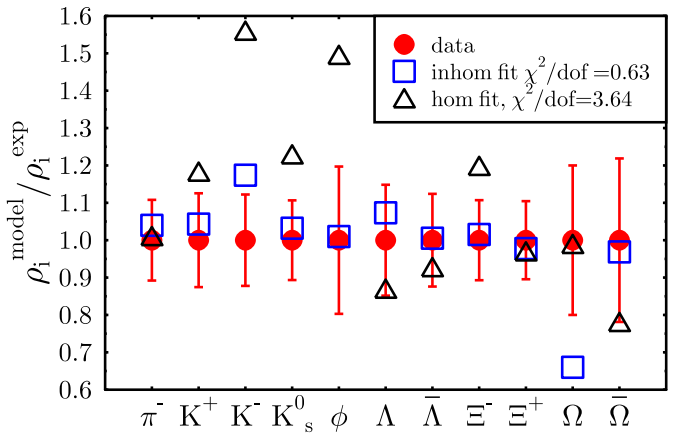}
\vspace{-0.6cm}
\caption{\label{ratios_lin}
Fitted particle densities divided by the corresponding measured
$4 \pi$ particle density (NA49 collaboration, Pb+Pb @ SPS 158 AGeV, compiled in
\protect{\cite{Becattini:2003wp}}) 
for the homogeneous 
($\delta T = \delta \mu = 0$) and the inhomogeneous fit 
($\delta T, \delta \mu$ free parameters). }
\end{figure}

The resulting particle ratios for the different fits are compared to 
the experimental $4 \pi$ data in figures \ref{ratios_log} and \ref{ratios_lin}.
As can be seen especially in figure \ref{ratios_lin},  
significant improvement compared to the homogeneous
freeze-out model is obtained for the Kaons, the mulit-strange baryons, but
also for the $\phi$, which couples only to temperature fluctuations.

\begin{figure}[h]
\vspace{-3.5cm}
\includegraphics[width=10cm]{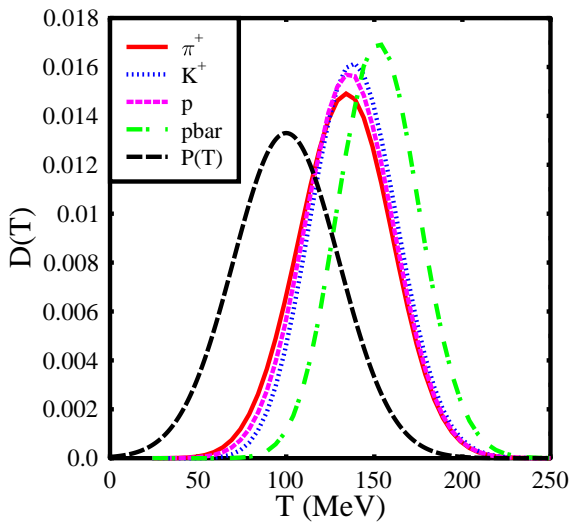}
\vspace{-1cm}
\caption{\label{ptdist}
Relative particle density versus temperature for 
$\pi, K, p$ and $\bar{p}$ which are obtained from folding the 
underlying Gaussian $(T,\mu)$-freeze-out distribution (dashed line) with 
the ideal-gas particle density distribution. }
\end{figure}

\begin{figure}[h]
\vspace{-3.5cm}
\includegraphics[width=10cm]{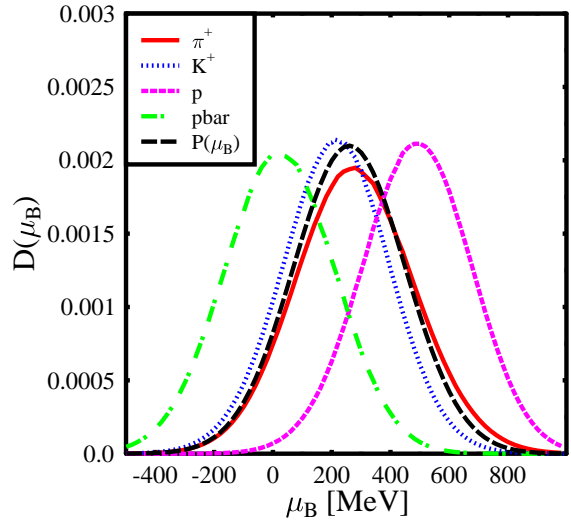}
\vspace{-1cm}
\caption{\label{pmudist}
Relative particle density versus chemical potential for 
$\pi, K, p$ and $\bar{p}$.}
\end{figure}

As can be seen from table \ref{tab1}, 
the inhomogeneous fits return significantly lower
mean temperature $\overline{T}$. However, these do {\em not} correspond to   
the ``mean'' emission temperature of the particles. 
The actual particle emission distribution is obtained by folding 
the 
assumed Gaussian $(T,\mu)$-freeze-out distribution 
(dashed line in Fig. \ref{ptdist} and \ref{pmudist}) with the ideal gas density
distribution for  
a given particle species.
The resulting normalized probability distributions read: 
\bea
& & D_i(T,\overline{T},\overline{\mu}_B, \delta T,\delta\mu_B)
= \\
& &  \frac {P(T;\overline{T},\delta T) 
\int\limits_{-\infty}^\infty 
d\mu_B \; P(\mu_B;
\overline{\mu}_B,\delta\mu_B)~\rho_i (T,\mu_B)~} {\overline{\rho}_i\; (\overline{T},\overline{\mu}_B, \delta T,\delta\mu_B)}. \nonumber
\eea
\bea
& & D_{i}(\mu_B,\overline{T},\overline{\mu}_B, \delta T,\delta\mu_B)
= \\
& &  \frac { P(\mu_B;
\overline{\mu}_B,\delta\mu_B)
\int\limits_{-\infty}^\infty 
d\mu_B \;P(T;\overline{T},\delta T)~\rho_i (T,\mu_B)~} {\overline{\rho}_i\; (\overline{T},\overline{\mu}_B, \delta T,\delta\mu_B)}. \nonumber
\eea
They are shown 
in figure \ref{ptdist} and \ref{pmudist}, respectively.
As expected from the temperature distribution we observe that the 
particle emission distributions are shifted towards higher temperatures.
How much the distribution is shifted depends on the mass and the chemical 
potential of the corresponding particle species. The  
probability distributions of the different particle species
versus chemical potential are shifted to larger chemical potentials for the particles and to lower chemical potentials for the antiparticles.

Thus, from the finite widths of the Gaussian, different particle emission 
distributions, with different peaks for different particle species, result.
The corresponding means of the distributions can be evaluated as
\bea
<T>_i &=& \int dT\; T\;  D_i(T,\overline{T},\overline{\mu}_B, \delta T,\delta\mu_B) ,\\
<\mu_B>_i &=& \int d\mu_B\; \mu_B\; 
D_{i}(\mu_B,\overline{T},\overline{\mu}_B, \delta T,\delta\mu_B).
\eea

The resulting means of the distributions for our analysis of the SPS
158 AGeV data 
are shown in table \ref{tab2}.

\begin{table}[ht]
\begin{center}
\begin{small}
\begin{tabular}{|c|cccccc|} \hline
SPS 158  & $p$ & $\bar{p} $ & $K^+$ & $K^-$ & $\Omega$ & $\bar{\Omega}$ \\ \hline
$<T>$ [MeV] (mid)& 157 & 170 &  152 & 150 & 164 & 180 \\ 
$<\mu_B>$ [MeV] (mid)& 268 & 191 & 237 & 222 & 234 & 225 \\
\hline
$<T>$ [MeV] ($4\pi$)& 136 & 153 &  140 & 139 & 151 & 165 \\ 
$<\mu_B>$ [MeV] ($4\pi$)& 487 & 22 & 306 & 213 & 277 & 206 \\
\hline 
\end{tabular}
\end{small}\end{center}
\vspace{-0.5cm}
\caption{\label{tab2}Mean temperature and chemical potential 
of various particle species for the inhomogeneous freeze-out.}
\end{table}
Again it can be seen that the antiparticles are mainly emitted from 
regions with small chemical potentials while the particles mainly 
originate from high chemical potential regions. The mean emission temperatures 
are in the range of 150-180 MeV for the midrapidity data and 
of 136-165 for the $4 \pi$ data. One sees that the antiparticles 
emerge in general from hotter regions    
than the particles. 

In conclusion, we find that 
allowing for inhomogeneities in the freeze-out temperature and 
chemical potential in an ideal gas description of particle production
in heavy ion collisions, significantly improves (lower $\chi^2/dof$) 
the description of experimental data at SPS 158 AGeV. 
It follows 
that the bulk of the particles originates from 
different density and temperature 
regions than the corresponding anti-particles. 
Hence, our results suggest that the decoupling surface might
not be very well ``stirred''. 
Furthermore, inhomogeneities appear to 
cure some deficencies of homogeneous freeze-out models and 
they might 
represent a potential variable to connect the measured particle 
abundances to the course of the expansion of the system.
The investigation of lower SPS and RHIC energies within our model is
under way \cite{Dumitru05b}.
Furthermore, in a future comprehensive analysis, 
the inhomogeneities should be generated
within a dynamical description. 

{\centerline{\em{Acknowledgements}}}

It is a pleasure to thank 
the organizers of the ''XLIII International Winter Meeting On Nuclear Physics''
in Bormio for the opportunity to present our results and my collaborators Adrian
Dumitru and Licinio Portugal. Furthermore I like to thank Carsten
Greiner for very fruitful discussions.

\end{document}